\documentclass[useAMS,usenatbib,twocolumn
]{mnras}

\usepackage[usenames,dvipsnames]{xcolor}

\usepackage{amssymb, amsmath}
\usepackage{epsf}
\usepackage{bm}
\usepackage{ulem}

\usepackage{natbib}
\usepackage{graphicx}
\usepackage{cancel}

\def\la{\; \raise0.3ex\hbox{$<$\kern-0.75em\raise-1.1ex\hbox{$\sim$}}\;}
\def\ga{\;  \raise0.3ex\hbox{$>$\kern-0.75em\raise-1.1ex\hbox{$\sim$}}\;}

\title[Magnetic field evolution timescales in superconducting neutron stars]
{Magnetic field evolution timescales in superconducting neutron stars}

\author[M.~E.~Gusakov, E.~M.~Kantor, D.~D.~Ofengeim]
{M.~E.~Gusakov\thanks{gusakov@astro.ioffe.ru},
E.~M.~Kantor,
D.~D.~Ofengeim
\\
Ioffe Institute, Polytekhnicheskaya 26, 194021 St.~Petersburg, Russia
}

\begin{document}

\newcommand{\PLeg}{\hat{\bf P}}
\newcommand{\mg}[1]{{\textcolor{orange}{\sf{#1}} }}
\newcommand{\ek}[1]{{\textcolor{RubineRed}{\sf{#1}} }}
\newcommand{\ekrm}[1]{\textcolor{RubineRed}{\sout{#1}}}
\newcommand{\dd}[1]{\textcolor[rgb]{0.0,0.8,0.2}{#1}}
\newcommand{\dddel}[1]{\textcolor[rgb]{0.0,0.8,0.2}{\sout{#1}}}

\date{Accepted 2020 xxxx. Received 2020 xxxx;
in original form 2020 xxxx}

\pagerange{\pageref{firstpage}--\pageref{lastpage}} \pubyear{2020}

\maketitle

\label{firstpage}

\begin{abstract}
The self-consistent approach to the magnetic field evolution in neutron star cores,
developed 
recently, is generalised to the case of superfluid and superconducting neutron stars. Applying this approach to the cold matter of neutron star cores composed of neutrons, protons, electrons, and muons we find that, similarly to the case of normal matter, an arbitrary configuration of the magnetic field may result in generation of macroscopic particle velocities, strongly exceeding their diffusive (relative) velocities. This effect substantially accelerates evolution of the magnetic field in the stellar core. An hierarchy of timescales of such evolution at different stages of neutron star life is proposed and discussed. It is argued that the magnetic field in the core cannot be considered as frozen or vanishing and that its 
temporal evolution
should affect the observational properties of neutron stars.
\end{abstract}


\maketitle

\section{Introduction}

The magnetic field in neutron stars (NSs) varies in a very wide range, from $10^8-10^{10}\,\rm G$ in millisecond pulsars to $10^{11}-10^{13}\,\rm G$ in standard (young) radio pulsars and up to $\sim 10^{15}\,\rm G$ in magnetars (\citealt*{kaspi10,vigano_etal13,pv19}). What processes drive the magnetic-field evolution in NSs? Only by answering this question can one understand, how different NS classes (with very different observation characteristics) relate to each other.

To build a 
comprehensive
theory describing magnetic-field evolution (B-evolution) in NSs,
one needs to understand the detailed properties of superdense matter in NS interiors in different phases (e.g., solid crust or liquid core) for a wide range of temperatures and magnetic field strengths. Up until now most of the work has been focused 
on the NS crust when modeling B-evolution (e.g., \citealt*{jones88, su97, rg02,
hr04, pg07, kk12, vigano_etal13, gcr_etal13, gc14, gwh16, gh18, gp20, kk20}).
These papers 
assume
that either the magnetic field in the core is absent or the currents in the stellar core, which generate a non-zero core field, virtually do not evolve in time. Although these assumptions simplify the problem considerably (because crust physics is much better understood than the core one), they basically cannot be justified without analysing 
magnetic field in the core. 

The processes responsible for the B-evolution in the core were discussed, in particular, by \cite*{gr92,us99, hrv08,hrv10,bl16,papm17,crv17,og18,crv20,crt19} in application to the normal NSs, and by \cite*{kg00,kg01,jones06,gjs11,gagl15,eprgv16,papm17,dg17, papm17b, blb18} in application to superfluid and superconducting NSs. Meanwhile, the detailed numerical modelling of the B-evolution simultaneously in the crust and in the core, accounting for the superconductivity of protons was carried out in two recent papers: \cite{eprgv16,blb18}. The results of all these studies are rather contradictory and reflect complexity and intricate character of the problem. A detailed discussion of these works would have taken up too much space, so here we will limit ourselves to a few comments, expressing our view on the problem of B-evolution in {\it superconducting} NSs.

1. The timescales of B-evolution derived by \cite{jones06,blb18} and \cite{gagl15,eprgv16} differ by many orders of magnitude. As shown by \cite{gusakov19}, very fast B-evolution of \cite{jones06,blb18} is a result of the incorrect treatment of forces acting on flux tubes in the NS core.

2. The timescale of non-dissipative B-evolution, $\tau_{\rm cons}$, derived by \cite{gagl15} is much 
larger than the similar timescale predicted by \cite{dg17}. This discrepancy was 
analysed
by \cite{papm17b}, 
who correctly concluded that for superfluid and superconducting NS matter in the core, composed of neutrons, protons, and electrons (npe-matter) the estimate of \cite{dg17} is inapplicable. However, as we show in section \ref{simple}, the situation changes drastically if one accounts for muons in the NS core. In this case the estimate of $\tau_{\rm cons}$, derived by \cite{dg17}, becomes appropriate.

3. All the papers mentioned above (except for \citealt{gko17,og18,crv20}) and discussing B-evolution in NSs assumed that in magnetised NS cores macroscopic fluid velocity of the core matter as a whole vanishes or is strongly suppressed. Generally, this assumption is not correct. In the paper by \cite{gko17} it was shown that fluid velocity is fully determined by the magnetic field configuration in the star. \cite{gko17} has also formulated a general approach to self-consistent calculation of this velocity, while \cite{og18} has demonstrated that fluid velocity, generally, strongly exceeds diffusive (relative) particle velocities, which results in substantial acceleration of B-evolution in NSs. This important result has recently been confirmed in detailed numerical simulations of \cite{crv20}.

\cite{og18} and \cite{crv20} discussed normal NSs in application to magnetars.
Yet, most of the observed NSs are cold $T\lesssim 10^7$~K, and have standard magnetic fields $10^{11}-10^{13}$~G (\citealt{km16}). When describing B-evolution in such stars, one necessarily has to allow for the 
effects of superfluidity and superconductivity of baryons in their internal layers. The impact of these effects on the equations of B-evolution is dramatic (see\ Secs.\ \ref{main} and \ref{eq}). Accordingly, the goals of this paper are: (i) to generalise the approach to self-consistent B-evolution, developed by \cite{gko17,og18}, to the case of superfluid and superconducting NS matter; (ii) to find the macroscopic fluid velocity in the core within this realistic approach and (iii) to estimate the timescales of B-evolution on different stages of NS life.

The paper is organised as follows. Section \ref{main} discusses the approximations adopted in the paper. In section \ref{eq} we formulate the main equations governing the B-evolution. The scheme of the solution to these equations, which is a direct generalisation of the approach developed by \cite{gko17,og18}, is introduced in section \ref{scheme}. Section \ref{simple} provides a traditional estimate of the conservative timescale, which would determine the non-dissipative B-evolution due to the tension and buoyancy forces in the absence of macroscopic and diffusive particle velocities. In section \ref{res} we present our main numerical results and propose estimates for the timescales of B-evolution in the NS core. We summarise and discuss our results in Section \ref{concl}. In Appendix we give some details on the derivation of the conservative timescale of B-evolution in $npe$-matter.

\section{Main assumptions}
\label{main}

We consider a model of cold NS with standard magnetic field. All equations below are written for ${\rm npe\mu}$-matter of NS cores, composed of neutrons (n), protons (p), electrons (e), and muons ($\mu$).%
%
\footnote{We added muons because for superfluid ${\rm npe}$-matter magnetic field cannot be arbitrary, but is constrained by an analogue of Grad-Shafranov equation (\citealt*{lander13,hw13}). Note that ${\rm npe}$-matter occupies only $\sim 400$~m of NS core in the model that we use (see below).}
%
We assume that the temperature $T$ is sufficiently low, so that neutrons (protons) can be considered as completely superfluid (superconducting). The effects of neutron-proton entrainment are neglected for simplicity (the off-diagonal element of entrainment matrix is set to zero, $\rho_{\rm np}=0$). Further, we consider a non-rotating NS with axial-symmetric magnetic field, ${\pmb B}={\pmb B}(r,\theta)$ ($r$ and $\theta$ are the radial coordinate and polar angle, respectively). Finally, we assume that protons form type-II superconductor, hence the magnetic field is confined to Abrikosov vortices and transported with the vortex velocity ${\pmb V}_{\rm L}$ (\citealt{kg01,gd16,blb18}), 
\begin{equation}
\frac{\partial {\pmb B}}{\partial t}
={\pmb \nabla} \times ({\pmb V}_{\rm L}\times {\pmb B}).
\label{Bevol}
\end{equation}
Equation (\ref{Bevol}) is simply Faraday's law with the electric field of the form ${\pmb E}=-{\pmb V}_{\rm L}\times {\pmb B}/c -{\pmb \nabla}\varphi_E$, where $\varphi_E$ is the electrostatic potential, which can be determined from the dynamic equations presented below, and $c$ is the speed of light. This electric field is obtained by Lorentz transformation of the magnetic field from the frame in which ${\pmb V}_{\rm L}=0$ to the laboratory frame.

\section{Basic equations}
\label{eq}

We shall be interested in the {\it quasistationary} evolution of the magnetic field (\citealt{gr92,gko17,og18}). We assume that, in the absence of magnetic field, an NS is spherically symmetric and is in full thermodynamic equilibrium. All deviations from the diffusion and beta-equilibrium are small and caused exclusively by the magnetic field; during the quasistationary evolution the star is in hydrostatic equilibrium, which is satisfied to a very high precision, because the corresponding timescale is much larger than the Alfven timescale (e.g., \citealt{gko17}). In this case, the dynamic equations governing the evolution can be substantially simplified (\citealt{gr92,gko17,og18}). Namely, one can neglect inertial terms in the Euler-like equations for particle species $j={\rm n}$, ${\rm p}$, ${\rm e}$, $\mu$ and, in addition, neglect time derivatives in all continuity equations (\citealt{og18}). For nonsuperfluid npe-matter the corresponding equations have been studied in detail, e.g., by \cite{gr92,papm17,og18}. For superfluid and superconducting npe-matter these equations were formulated by \cite{gas11,kg17,papm17b} and are implicitly contained in the paper by \cite{gd16}. An extension of these equations to include muons is straightforward. In what follows the equation of state (EOS) is assumed to be relativistic, but the effects of general relativity are disregarded for simplicity. The whole system consists of:

(1) Continuity equations for all particle species (here and hereafter the index $i$ refers to nucleons, $i={\rm n}$, ${\rm p}$ and the index $l$ refers to leptons, $l={\rm e}$, $\mu$)
\begin{equation}
{\pmb \nabla}\cdot (n_i {\pmb V}_{{\rm s}i}) =0, \quad \quad
{\pmb \nabla}\cdot (n_l {\pmb u}_{l}) =0,
\label{cont}
\end{equation}
where ${\pmb V}_{{\rm s}i}$ and ${\pmb u}_l$ are the velocity of nucleon species $i$ and lepton species $l$, respectively; $n_j$ is the number density for particle species $j={\rm n}$, $\rm p$, $\rm e$, $\mu$. We omit the sources due to beta-processes in these equations; they are negligible in cold superfluid and superconducting interiors of NSs.

(2) Euler-like equations
\begin{align}
&
n_{\rm p} {\pmb \nabla} \mu_{\rm p}^\infty = 
{\pmb { \mathcal F}}_{\rm b+t}
-{\pmb {\mathcal F}}_{\rm pe}
-{\pmb {\mathcal F}}_{{\rm p}\mu}
+e_{\rm p} n_{\rm p} \, 
\left(
{\pmb E}+\frac{{\pmb V}_{\rm sp}}{c}\times {\pmb B}
\right),
&
\label{euler_p}\\
&
n_{\rm n} {\pmb \nabla} \mu_{\rm n}^\infty = 0,
&
\label{euler_n}\\
&
n_{\rm e} {\pmb \nabla} \mu_{\rm e}^\infty = 
{\pmb {\mathcal F}}_{\rm \mu e}+
{\pmb {\mathcal F}}_{\rm pe}+
e_{\rm e} n_{\rm e} \, \left(
{\pmb E}+\frac{{\pmb u}_{\rm e}}{c}\times {\pmb B}
\right),
&
\label{euler_e}\\
&
n_{\rm \mu} {\pmb \nabla} \mu_{\rm \mu}^\infty = 
-{\pmb {\mathcal F}}_{\rm \mu e}+
{\pmb {\mathcal F}}_{\rm p \mu }+
e_{\rm \mu} n_{\rm \mu} \, \left(
{\pmb E}+\frac{{\pmb u}_{\rm \mu}}{c}\times {\pmb B}
\right). 
&
\label{euler_mu}
\end{align}
Here $\mu_j^\infty =\mu_j {\rm e}^{\phi/c^2}$; $\phi$ is the gravitational potential; $\mu_j$ and $e_j$ are, respectively, the relativistic chemical potential and electric charge for particle species $j$; ${\pmb {\mathcal F}}_{\rm \mu e}=J_{\rm \mu e} ({\pmb u}_{\mu}-{\pmb u}_{\rm e})$ is the friction force between electrons and muons; $J_{\rm \mu e}$ is the corresponding  momentum transfer rate taken from \cite{shternin08,dgs20} 
\footnote{\label{foot2} $J_{\rm \mu e}$ in \cite{shternin08,dgs20} is calculated under the assumptions that protons are non-superconducting. Proton superconductivity may affect $J_{\rm \mu e}$.}. 
Further, ${\pmb {\mathcal F}}_{\rm b+t}=({\nabla}\times {\pmb H}_{\rm c1})\times {\pmb B}/(4 \pi)$ is the combined buoyancy and tension forces which act on vortices (\citealt{dg17}); ${\pmb H}_{\rm c1}=H_{\rm c1}(r) {\pmb e}_B$, where ${\pmb e}_B$ is a unit vector along ${\pmb B}$, $H_{\rm c1}(r)\sim (10^{14}-10^{15})$~G is the first critical magnetic field. It is determined by EOS and critical temperature of proton superfluidity onset, $T_{\rm cp}$, and equals (\citealt{ll80}) $H_{\rm c1}=4\pi \epsilon_{\rm p}/\Phi_0$, where $\Phi_0$ is a quantum of magnetic field flux, $\epsilon_{\rm p}$ is energy of proton vortex per unit length. For neutron star matter, where proton coherence length is typically of the order of London penetration depth, $\epsilon_{\rm p}$ was calculated by \cite{mendell91a}.
Finally, ${\pmb {\mathcal F}}_{{\rm p}l}=\mathcal{D}_l \, {\pmb e}_{B}\times[{\pmb e}_{B}\times ({\pmb u}_{l}-{\pmb V}_{\rm L})]$ is the dissipative force appearing due to scattering of electrons (muons) off the vortex magnetic field, where the coefficients $\mathcal{D}_l\propto B$ depend on density but are independent of temperature in the limit $T\ll T_{\rm cp}$. The non-dissipative part of the force appearing due to scattering of electrons (muons) off the vortex magnetic field is already taken into account in equations (\ref{euler_e}) and (\ref{euler_mu}): in the limit $T\ll T_{\rm cp}$ it is the Lorentz force. Indeed, using the expression for ${\pmb E}$ from section \ref{main}, one may, e.g., write: $e_{l} n_{l} \, \left({\pmb E}+{\pmb u_{l}}\times {\pmb B}/c \right)
= \mathcal{D}'_l ({\pmb u_{l}}-{\pmb V}_{\rm L})\times {\pmb e}_{B} - e_{l} n_{l} {\pmb \nabla} \varphi_E$, where $\mathcal{D}'_l=e_{l} n_{l} B/c$ is one of the coefficients calculated by \cite{gusakov19} and the last term arises because of spatial inhomogeneity of the system (non-vanishing macroscopic density gradients). The discussed coefficients are presented in Fig. \ref{Fig:D} (see also \citealt{gusakov19}).

(3) Charge neutrality condition, $e_{\rm p} n_{\rm p}+e_{\rm e}n_{\rm e}+e_\mu n_{\mu}=0$, combined with the screening condition (\citealt{jones91,gas11,gd16})
\begin{equation}
e_{\rm p} n_{\rm p} {\pmb V}_{{\rm sp}} +e_{\rm e} n_{\rm e} {\pmb u}_{\rm e}+e_{\rm \mu} n_{\rm \mu} {\pmb u}_{\rm \mu}=0.
\label{screening}
\end{equation}

(4) The condition expressing the fact that sum of forces acting on vortices must vanish (\citealt{gusakov19})
\begin{equation}
{\pmb {\mathcal F}}_{\rm Magnus}
-{\pmb {\mathcal F}}_{\rm pe}-{\pmb {\mathcal F}}_{\rm p\mu}+{\pmb {\mathcal F}}_{\rm b+t}=0,
\label{vortcond}
\end{equation}
where ${\pmb {\mathcal F}}_{\rm Magnus}
=-(e_{\rm p} n_{\rm p}/c) \, {\pmb B}\times({\pmb V}_{\rm sp}-{\pmb V}_{\rm L})$ 
is the Magnus force (\citealt{sonin87}).
Expressing ${\pmb {\mathcal F}}_{\rm b+t}-{\pmb {\mathcal F}}_{\rm pe}-{\pmb {\mathcal F}}_{\rm p\mu}$ from equation~(\ref{vortcond}) and substituting it into (\ref{euler_p}), gives ${\pmb \nabla} \mu_{\rm p}^\infty=-e_{\rm p}{\pmb \nabla} \varphi_{E}$. Equations (\ref{euler_p})--(\ref{screening}) allow one to derive the total force balance equation in the form
\begin{equation}
n_{\rm e} {\pmb \nabla} d_{\rm e\mu}^\infty + n_{\rm p} {\pmb \nabla} \Delta\mu_\mu^\infty = {\pmb {\mathcal{F}}_{\rm b+t}},
\label{tot}
\end{equation}
where $d_{\rm e\mu}^\infty=\mu_{\rm e}^\infty-\mu_\mu^\infty$ and $\Delta \mu_{\mu}^\infty \equiv \mu_{\mu}^\infty+\mu_{\rm p}^\infty-\mu_{\rm n}^\infty$ are the (redshifted) imbalances of chemical potentials, which are generated by the magnetic field. Without the magnetic field the matter is in beta-equilibrium and $d_{\rm e\mu}^\infty=\Delta \mu_{\mu}^\infty=0$.

\section{Scheme of the solution}
\label{scheme}

Our aim here is to calculate all the particle velocities for a given magnetic field configuration and hence to estimate the corresponding timescale of the B-evolution. Since the algorithm of the velocity finding is very similar to that discussed in detail by \cite{gko17,og18}, we only briefly describe its main steps in what follows. Below we introduce the operator $\PLeg_m$, that extracts $m$-th Legendre component of its argument, i.e., $\PLeg_m(\cdot)=(2m+1)/2 \, \int_0^\pi (\cdot) P_m({\rm cos}\theta) {\rm sin}\theta {\rm d}\theta$, where $P_m({\rm cos}\theta)$ is the Legendre polynomial of degree $m$.

($i$) We assume that the magnetic field ${\pmb B}(r,\theta)$ is specified.

($ii$) It perturbs the redshifted chemical potential imbalances $d_{\rm e\mu}^\infty$ and $\Delta \mu_\mu^\infty$. The Legendre components $m\geq 1$ of these imbalances can be found from equation (\ref{tot}):
\begin{equation}
\label{eq:DmudmuHighL}
\begin{pmatrix}
\PLeg_m d_{\rm e\mu}^\infty \\
\PLeg_m\Delta\mu_{\mu}^\infty
\end{pmatrix}
=
\begin{pmatrix}
n_e' & n_p' \\
n_e & n_p
\end{pmatrix}^{-1} 
\begin{pmatrix}
\left[ r \PLeg_m R \right]' 
- \PLeg_m \mathcal{F}_{{\rm b+t}, r}\\
r \PLeg_m R
\end{pmatrix}, 
\end{equation}
where the function $R(r,\theta)\equiv \int_0^\theta \mathcal{F}_{\rm b+t,\theta}{\rm d}\theta$; and prime $'$ means $d/dr$. Note that components  $\PLeg_0 d_{\rm e\mu}^\infty$ and $\PLeg_0 \Delta \mu_\mu^\infty$ remain undefined and will be found at step 5.

\begin{figure}
\begin{center}
	\includegraphics[width=\columnwidth]{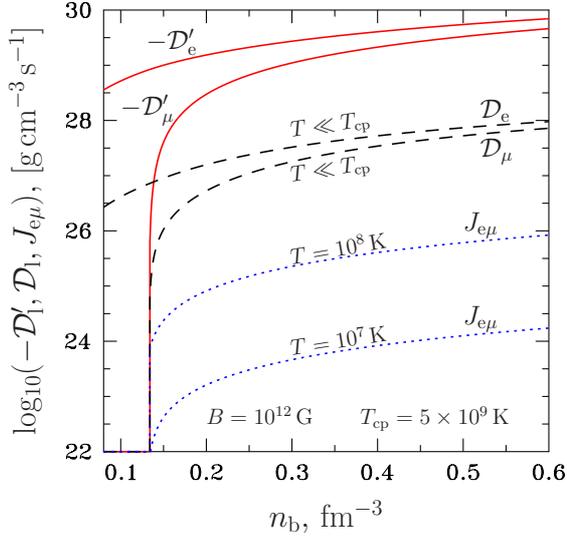}
	\caption{Coefficients $\mathcal{D}_l$, $\mathcal{D}'_l$ ($l={\rm e}, \, \mu$), and $J_{\rm \mu e}$ versus baryon number density $n_{\rm b}$ for $T_{\rm cp}=5\times 10^9\,\rm K$ and equation of state from section \ref{res}. Coefficients $\mathcal{D}_l$ and $\mathcal{D}'_l$ are temperature-independent in the limit $T \ll T_{\rm cp}$. $J_{\rm \mu e}$ depends on temperature and is presented for $T=10^7\,\rm K$ and $T=10^8\,\rm K$.}
	\label{Fig:D} 
	\end{center}
\end{figure}

($iii$) Determine the relative muon-electron velocity, $\Delta {\pmb u}_{\rm \mu e}\equiv {\pmb u}_\mu-{\pmb u}_{\rm e}$, from equations of section \ref{eq}. The resulting expression is, generally, very lengthy, but it can be simplified using the fact that $\mathcal{D}_l \gg J_{\rm \mu e}$, which is true for cold NSs ($T \sim 10^7\,\rm K$) at not too low magnetic fields $B\gtrsim 10^9$~G (see figure~\ref{Fig:D}). Taking into account also that
\footnote{For muons this condition is not satisfied in the immediate vicinity of the muon onset density.}
$\mathcal{D}_l \ll |\mathcal{D}'_l|$
(see \citealt{gusakov19} and figure~\ref{Fig:D}), one can approximately write:
\begin{equation}
\Delta {\pmb u}_{\rm \mu e}\approx
\mathcal{K}\, {\pmb \nabla}\tilde{d}_{\rm e\mu}^\infty
+ \frac{c}{e_{\rm p} B} {\pmb e}_{B}\times {\pmb \nabla}\tilde{d}_{\rm e\mu}^\infty
+ \frac{n_{\rm e}n_\mu}{n_{\rm p} J_{\rm \mu e}} \, 
{\pmb e}_{B} ({\pmb e}_{B} \cdot {\pmb \nabla}\tilde{d}_{\rm e\mu}^\infty),
\label{umuMinue}
\end{equation}
where $\mathcal{K}=(\mathcal{D}_\mu n_{\rm e}^2+\mathcal{D}_{\rm e}n_\mu^2)/(n_{\rm p} \mathcal{D}'_{\rm \mu} \mathcal{D}'_{\rm e})$, 
\begin{eqnarray}
{\pmb \nabla}\tilde{d}_{\rm e\mu}^\infty \approx {\pmb \nabla}d_{\rm e\mu}^\infty+\frac{\mathcal{D}_{\rm e}n_{\rm \mu}-\mathcal{D}_{\rm \mu}n_e}{n_{\rm e} n_{\rm \mu}}{\pmb e}_B\times({\pmb e}_B\times \delta {\pmb V}_{\rm sp}),
\label{dtilde}
\end{eqnarray} 
and $\delta {\pmb V}_{\rm sp}\equiv \frac{c}{4\pi e_{\rm p}n_{\rm p}}{\pmb\nabla}\times {\pmb H}_{\rm c1}$. For an arbitrary configuration of the magnetic field ${\pmb B}$, the second term in (\ref{dtilde}) is generally $\mathcal{D}_{l}/\mathcal{D}'_{l}$ times smaller than the first one. The coefficients in (\ref{umuMinue}) are ordered as $n_{\rm e}n_\mu/(n_{\rm p}J_{\rm \mu e})\gg c/(e_{\rm p}B)\gg \mathcal{K}$. 
Working in the same approximation as when deriving (\ref{umuMinue}), the vortex velocity ${\pmb V}_{\rm L}$ can be determined from (\ref{vortcond}) and presented as
\begin{multline}
{\pmb V}_{\rm L} \approx {\pmb V}_{\rm sp}+\frac{c {\pmb \nabla}\times {\pmb H}_{\rm c1}}{4 \pi e_{\rm p}n_{\rm p}} + \mathcal{K}_1 {\pmb e}_{B}\times ({\pmb e}_{B} \times {\pmb \nabla}{\tilde d}_{\rm e\mu}^\infty) \\ 
+ \mathcal{K}_2 \frac{{\pmb \nabla}\times {\pmb H}_{\rm c1}}{4 \pi}  \times {\pmb B},
\label{Vl}
\end{multline}
where 
$\mathcal{K}_1 
= (\mathcal{D}_\mu n_{\rm e}-\mathcal{D}_{\rm e}n_\mu)/(\mathcal{D}'_{\rm e}+\mathcal{D}'_{\mu})^2$ 
and 
$\mathcal{K}_2 
= (\mathcal{D}_{\rm e} +\mathcal{D}_{\rm \mu})/(\mathcal{D}'_{\rm e}+\mathcal{D}'_{\mu})^2$.
Note that the approximate formulas (\ref{umuMinue})---(\ref{Vl}) are not applicable near the muon onset density.

($iv$) Express poloidal components of one of the velocities, e.g., ${\pmb u}_{e}$, through $\Delta {\pmb u}_{\rm \mu e}$ using the continuity equations (\ref{cont}):
\begin{align}
&
u_{{\rm e}, r}=-\frac{{\pmb \nabla}\cdot 
	[n_{\rm \mu}\Delta {\pmb u}_{\rm \mu e}]}
{n_{\rm e} 
	\nabla \left( n_{\mu}/n_{\rm e} \right)},
&
\label{umur}\\
&
u_{{\rm e}, \theta}=-\frac{1}{{\rm sin}\theta}\left[\frac{1}{n_{\rm e} r } \frac{\partial}{\partial r}
\left(
r^2 n_{\rm e}\int_{0}^\theta u_{{\rm e}, r} {\rm sin}\widetilde{\theta} {\rm d}\widetilde{\theta}
\right)+\xi (r)\right],
&
\label{umutheta}
\end{align}
where $\xi(r)$ is some function. 

An interesting feature of the solution (\ref{umur})--(\ref{umutheta}) is that in the axisymmetric problem the poloidal components $u_{{\rm e}, r}$ and $u_{{\rm e},\theta}$
appear to be independent of the toroidal component of the relative velocity $\Delta {\pmb u}_{\mu {\rm e}}$.

The toroidal component of ${\pmb u}_{\rm e}$ does not appear in the continuity equations and should be found from a separate requirement 
\begin{equation}
\frac{\partial {F}_{{\rm b+t},\phi}}{\partial t}=0, 
\label{uphi}
\end{equation}
analogous to that formulated for the non-superfluid NS matter (\citealt{gko17,og18}). Here we do not attempt solving this equation.

Note that, once $\Delta {\pmb u}_{\rm \mu e}$ and ${\pmb u}_e$ are calculated, it is easy to find ${\pmb u}_\mu = {\pmb u}_e + \Delta {\pmb u}_{\rm\mu e}$ and ${\pmb V}_{\rm sp} = {\pmb u}_{\rm e} + n_\mu/n_{\rm p}\Delta {\pmb u}_{\rm\mu e}$ using equation~(\ref{screening}).

($v$) The velocity components $u_{{\rm e}, r}$, $u_{{\rm e}, \theta}$ and their derivatives should be finite everywhere in the core. Moreover, due to the axial symmetry of the problem, $u_{{\rm e}, \theta}$ must vanish on the symmetry axis at arbitrary $r$: $u_{{\rm e}, \theta}(r,0)=0$, $u_{{\rm e}, \theta}(r,\pi)=0$. Accounting for these conditions, one immediately obtains (see equation~\ref{umutheta} and \citealt{og18}):
\begin{align}
\xi (r) &= 0,
\\
\int_{0}^\pi u_{{\rm e}, r} {\rm sin} \widetilde{\theta}\,  {\rm d} \widetilde{\theta} &= 0.
\label{1}
\end{align}
Note that, in view of equation~(\ref{umur}), $u_{{\rm e}, r}$ depends on the difference $\Delta {\pmb u}_{\rm \mu e}$. The latter difference, in turn, depends on the imbalance $d_{\rm e\mu}^\infty$ (see equation~\ref{umuMinue}), which can be expanded in Legendre polynomials as: $d_{\rm e\mu}^\infty(r,\theta)=\PLeg_0 d_{\rm e\mu}^\infty +\sum_{m=1}^\infty (\PLeg_m d_{\rm e\mu}^\infty)\, P_m({\rm cos}\theta)$. All components in this expansion except for $m=0$ are known (see equation \ref{eq:DmudmuHighL}), thus the formula~(\ref{1}) should be considered as a differential equation for the zero'th component $\PLeg_0 d_{\rm e\mu}^\infty$. This is a second-order inhomogeneous differential equation. It should be solved with the boundary conditions $\PLeg_0 d_{\rm e\mu}^\infty|_{r=0}=0$, $(\PLeg_0 d_{\rm e\mu}^\infty)'|_{r=0}=0$ (see \citealt{og18} for a discussion of similar conditions); the last one is a necessary requirement for the finiteness of $u_{{\rm e},r}$ at $r\rightarrow 0$. Solution to this equation together with equation~(\ref{eq:DmudmuHighL}) allows one to determine $d_{\rm e\mu}^\infty$ and hence $\Delta {\pmb u}_{\rm \mu e}$, poloidal components of ${\pmb u}_{\rm e}$, ${\pmb V}_{\rm L}$, as well as poloidal components of all other velocities (see equations \ref{umuMinue}---\ref{umutheta}).

\begin{figure*}
	\includegraphics[width=0.9\textwidth]{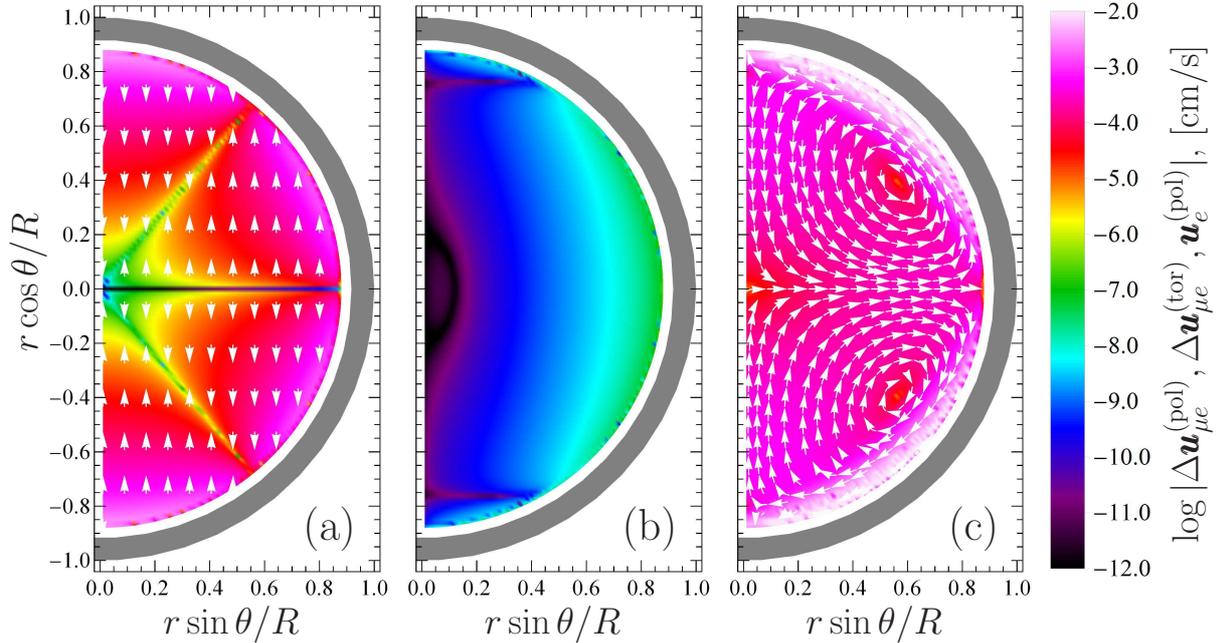}
	\caption{Panels (a) and (b) present, respectively, poloidal and toroidal components of $\Delta{\pmb u}_{\mu e}$. Panel (c) presents poloidal component of ${\pmb u}_{\rm e}$. The homogeneous magnetic field is directed upwards. Different colours correspond to different values of the common logarithm (${\rm log}_{10}$) of the corresponding velocity field; arrows in panels (a) and (c) show the direction of the poloidal velocity component. Due to numerical problems we do not carry out the calculations in the vicinity of muon threshold (white region). Crust is shaded grey.}
\label{fig1}
\end{figure*}

\section{Traditional estimate}
\label{simple}

Before we proceed with the numerical results, let us present a `naive' estimate for the typical timescale $\tau$ of the B-evolution, which can often be found in the literature (e.g., \citealt{gagl15,dg17}). According to Eq.\ (\ref{Bevol}) $\tau = B/|{\pmb \nabla}\times ({\pmb V}_{\rm L}\times {\pmb B})| \sim L/V_{\rm L}$, where $L$ is the typical lengthscale of ${\pmb B}$ and ${\pmb V}_{\rm L}$ variation in the star, $L\sim 10^5-10^6$~cm. We can find ${\pmb V}_{\rm L}$ from (\ref{Vl}). Usually, when estimating ${\pmb V}_{\rm L}$ the first term (the velocity ${\pmb V}_{\rm sp}$) is neglected (as we show in section~\ref{res} this approximation is incorrect). Taking also into account that the `dissipative' terms, $\propto \mathcal{K}_1$ and $\propto \mathcal{K}_2$, are much smaller than the second term 
\footnote{
This follows from the inequality $\mathcal{D}_l\ll\mathcal{D}_l'$ 
and the estimate ${\pmb \nabla} d_{\rm e\mu}^\infty \sim {\pmb {\mathcal{F}}_{\rm b+t}}/n_{\rm e}$ (see \ref{tot}).
},
${\pmb V}_{\rm L}$ can be written as ${\pmb V}_{\rm L} \sim c {\pmb \nabla}\times {\pmb H}_{\rm c1}/(4 \pi e_{\rm p} n_{\rm p})$,
and, using 
the definition of $\tau$, one gets
\begin{equation}
\tau \sim 3 \times 10^8 \, \frac{L_6^2  }{ H_{\rm c1, 15}}\,\frac{n_{\rm p}}{0.05\, \rm fm^{-3}}\,\,\, \ {\rm yr},
\label{tau1}
\end{equation}
where $L_6=L/(10^6 \ {\rm cm})$, $H_{\rm c1, 15}=H_{\rm c1}/(10^{15}\ {\rm G})$.
Our estimate (\ref{tau1}) is consistent with the similar estimate for the conservative timescale $\tau_{\rm cons}$ in the paper by \cite{dg17}, see 
appendix in the end of this paper
for the further discussion and comparison with the results of \cite{papm17}.

\section{Results}
\label{res}

We applied the scheme from section~\ref{scheme} to calculate $\Delta {\pmb u}_{\rm \mu e}$ and the poloidal component of the velocity  ${\pmb u}_{\rm e}$. Knowing these quantities, it is easy to find the poloidal components of all other velocities. In our calculations we used the same microphysical input as in \cite{og18}. Namely, we used equation of state HHJ (\citealt*{hhj99}) in the NS core. All results are obtained for the star with the mass $M=1.4 M_\odot$, which has the radius $R=12.1$~km and the core radius $R_{\rm core}=11.1$~km. Muons appear at $R_\mu=10.7$~km. In what follows calculations are performed only in the inner core where muons are present (at $r<R_\mu$). To simplify the calculations we consider a {\it homogeneous} magnetic field model. Note that, in superconducting matter a homogeneous magnetic field is not a force-free field (${\pmb {\mathcal F}}_{\rm b+t}\neq0$), because of the action of buoyancy force on the flux tubes (\citealt{dg17}). Below we discuss how this choice of the field configuration may affect our conclusions. All results below are derived for $B=10^{12}$~G, internal temperature $T=10^7$~K, and 
$T_{\rm cp}=5\times 10^9\,\rm K$.

Figure~\ref{fig1} presents the results for the poloidal and toroidal components of $\Delta {\pmb u}_{\rm \mu e}$ (panels a and b, respectively), calculated with formulas of section~\ref{eq} not making any approximations like during derivation of equation~(\ref{umuMinue}). The value of 
${\rm log}_{10}|\Delta {\pmb u}_{\rm \mu e}^{(\rm pol)}|$ and ${\rm log}_{10}|\Delta {\pmb u}_{\rm \mu e}^{(\rm tor)}|$
is indicated by different colours, arrows in panel (a) show the direction of the corresponding poloidal component. The homogeneous magnetic field is directed upwards. As one can see, the poloidal component of $\Delta {\pmb u}_{\rm \mu e}$ is 
almost
collinear with ${\pmb B}$ and strongly exceeds the toroidal component. Mathematically this can be understood if we recall that for the chosen ${\pmb B}$ and $T$ $|D_l'|\gg D_l \gg J_{\rm \mu e}$ (see section~\ref{scheme} and figure~\ref{Fig:D}), and, as a result, the third term in the r.h.s.\ of equation~(\ref{umuMinue}), which is collinear to ${\pmb e}_{B}$, strongly exceeds the other terms in this equation. Physically, the component of $\Delta {\pmb u}_{\rm \mu e}$ along ${\pmb B}$ appears to be large because electrons and muons moving along ${\pmb e}_B$ do not scatter off the magnetic field of flux tubes, and can only rub against each other; such friction, however, is small ($\sim J_{\rm \mu e}$).

Very rough estimate with equations~(\ref{tot}) and~(\ref{umuMinue}) under the assumption $n_{\rm e}\sim n_\mu$ 
gives for our field configuration
\begin{align}
&
\Delta u_{\rm \mu e}^{({\rm pol})} \sim \frac{H_{\rm c1} B}{4 \pi L J_{\rm \mu e}}
\sim 10^{-4}  \, \frac{H_{\rm c1, 15} B_{12}}{L_6 T_7^\gamma}\ {\rm cm\ s^{-1}},
& 
\label{pol}\\
&
\Delta u_{\rm \mu e}^{({\rm tor})} \sim \frac{c H_{\rm c1}}{4 \pi e_{\rm p}L n_{\rm e}}
\sim 10^{-10} \, \frac{H_{\rm c1,15}}{L_6 }\, \frac{0.05\,\rm fm^{-3}}{n_{\rm e}} \ {\rm cm\ s^{-1}},
&
\label{tor}
\end{align}
where we take $\gamma=5/3$, as in \cite{shternin08,dgs20} (but see footnote \ref{foot2}).

Let us now look at figure~\ref{fig1}(c), showing poloidal component $u_{\rm e}^{({\rm pol})}$ of ${\pmb u}_{\rm e}$, calculated using equations~(\ref{umur}) and~(\ref{umutheta}). One can see that $u_{\rm e}^{({\rm pol})}$ is much larger than $\Delta u_{\rm \mu e}^{({\rm pol})}$. On average, $u_{\rm e}^{({\rm pol})} \sim C \, \Delta u_{\rm \mu e}^{({\rm pol})}$, with the coefficient $C \sim 10$. 
This result is not surprising. The point is $u_{\rm e}^{({\rm pol})}$, generally, depends on high (up to the fourth) spatial derivatives of ${\pmb H}_{\rm c1}$, ${\pmb B}$, and number densities $n_{\rm e}$ and $n_{\rm \mu}$. These derivatives are very large in the outer NS layers, especially near the muon onset density, that results in a big difference between $u_{\rm e}^{({\rm pol})}$ and $\Delta u_{\rm \mu e}^{({\rm pol})}$. A similar effect was discovered and studied in detail for non-superconducting NSs by \cite{og18}.

Since $u_{\rm e}^{({\rm pol})} \gg \Delta u_{\rm \mu e}^{({\rm pol})}$, while all other velocity differences can be expressed through each other and has the same order of magnitude (see section \ref{scheme}), one can conclude that all particle species and flux tubes move with approximately one and the same velocity $u_{\rm e}^{({\rm pol})}$. In particular, poloidal component of the vortex velocity  ${\pmb V}_{\rm L}^{({\rm pol})}\approx u_{\rm e}^{({\rm pol})} \sim C \, \Delta u_{\rm \mu e}^{({\rm pol})}\propto B/T^\gamma$. 
Using this estimate together with equation~(\ref{pol}), we find that the typical timescale of the B-evolution is
\begin{equation}
\tau_{\rm init} \sim \frac{L}{V_{\rm L}} \sim 
\frac{400}{C}
\frac{L_6^2 T_7^\gamma}{H_{\rm c1,15}B_{12}} \ {\rm yr}.
\label{tau2}
\end{equation}
and we remind that $C\sim 10$, as follows from our numerical results. This time is extremely short compared to the typical pulsar age $\sim 10^7$~yr (%
see, e.g., the ATNF pulsar catalogue, \citealt{ATNFref}\footnote{\texttt{www.atnf.csiro.au/people/pulsar/psrcat/}}; also 
\citealt{harding13,km16})%
.

Although we do not have a rigorous proof, it seems natural that $\tau_{\rm init}$ should be a typical timescale of the magnetic field rearrangement in order to decrease the huge term $\propto 1/J_{\rm \mu e}$ in equation~(\ref{umuMinue}). Such rearrangement tends to make this term comparable to other terms in this equation, which implies the condition
\begin{equation}
{\pmb e}_B \cdot {\pmb \nabla}\tilde{d}_{\rm e \mu}^\infty ={\pmb e}_B \cdot {\pmb \nabla}{d}_{\rm e \mu}^\infty\approx 0.
\label{cond}
\end{equation}
In other words, ${d}_{\rm e \mu}^\infty$ tends to be approximately constant along the magnetic field lines. Since, as we showed, ${\pmb \nabla}{d}_{\rm e \mu}^\infty$ is completely determined by the magnetic field, the condition (\ref{cond}) should be considered as indirect constraint on the possible configuration of the magnetic field in the superconducting ${\rm npe\mu}$-core. Note that for purely toroidal axisymmetric magnetic field such rearrangement does not occur, because then the condition (\ref{cond}) is satisfied automatically. We should also note 
that, since $\tau_{\rm init}$ is much smaller than the NS cooling timescale, $\tau_{\rm cool}$, 
for real NSs
the field rearrangement 
occurs on the cooling timescale, 
on which they approach the low-temperature regime
studied
in the paper (see also the discussion below in this section).

Here we come to the question what will be the typical timescale of the B-evolution {\it after} the magnetic field reaches the configuration satisfying the condition~(\ref{cond})? To get an impression of this timescale one can calculate $\Delta {u}_{\rm \mu e}^{({\rm pol})}$ and ${u}_{\rm e}^{({\rm pol})}$ (and hence $V_{\rm L}^{({\rm pol})}$) with equations~(\ref{umuMinue}), (\ref{umur}), and~(\ref{umutheta}), for the same homogeneous field configuration, but artificially replacing the huge last term in equation~(\ref{umuMinue}) with $c/(e_{\rm p}B) {\pmb e}_{B} ({\pmb e}_{B} \cdot {\pmb \nabla}{d}_{\rm e\mu}^\infty)$. 
It has the similar form, but its
absolute value is comparable to the second term in (\ref{umuMinue}).%
\footnote{
Note that, since the employed magnetic-field model (${\pmb B}={\rm const}$) is purely poloidal, the second term in (\ref{umuMinue}) is directed along ${\pmb e}_\varphi$ ($\varphi$ is the azimuthal angle) and thus does not contribute to ${u}_{\rm e}^{({\rm pol})}$ in the axisymmetric problem [see (\ref{umur}) and note that ${\pmb \nabla}\cdot [A(r,\theta) {\pmb e}_{\varphi}]=0$ for any function $A(r,\theta)$]. Nevertheless, for more realistic field models, possessing the toroidal component (see, e.g., \citealt{bs04}), this term will contribute to ${u}_{\rm e}^{({\rm pol})}$.
}
%
\begin{figure*}
	\includegraphics[width=0.9\textwidth]{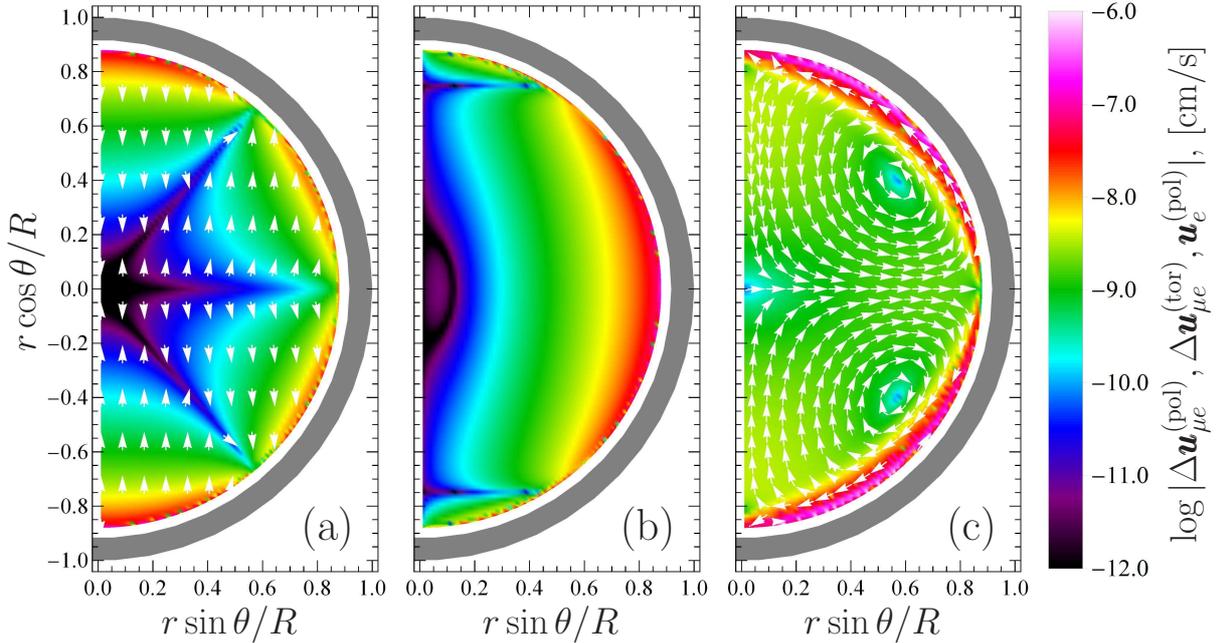}
	\caption{The same as figure~\ref{fig1}, but after the fast relaxation of the magnetic field is finished (see the text). 
	The toroidal component of $\Delta {\pmb u}_{\rm \mu e}$, shown in panel (b), is the same as in figure~\ref{fig1}(b), but is plotted using different colour scheme to facilitate comparison with panels (a) and (c).
}
\label{fig2}
\end{figure*}

The result of calculation is presented in figure~\ref{fig2}. Due to the same reasons as in the previous calculation (accounting for the last term in equation~\ref{umuMinue} in its original form), ${u}_{\rm e}^{({\rm pol})}$ appears to be $C \sim 10$ times larger than $\Delta {u}_{\rm \mu e}^{({\rm pol})}$. Keeping in mind that the first term, $\mathcal{K} {\pmb \nabla} {\tilde d}_{\rm e\mu}^\infty$, in the expression for $\Delta {\pmb u}_{\rm \mu e}$ is of the order of $\sim \mathcal{D}_l'/\mathcal{D}_l$, that is $\sim 100$ times smaller than the remaining terms in our artificial problem (see section \ref{scheme} and figure~\ref{Fig:D}), $u_{\rm e}^{({\rm pol})}$ (and $V_{\rm L}^{({\rm pol})}$) can be estimated as: 
\begin{multline}
u_{\rm e}^{({\rm pol})} \sim 
C \Delta {u}_{\rm \mu e}^{({\rm pol})} \sim 
C \frac{c |{\pmb \nabla} {d}_{\rm e \mu}^\infty|}{e_{\rm p} B} \sim C \frac{c H_{\rm c1}}{4 \pi e_{\rm p} L n_{\rm e}}
\\ 
\sim 10^{-10}C\, \frac{H_{\rm c1,15}}{L_6} \frac{0.05\,\text{fm}^{-3}}{n_{\rm e}} \text{~cm~s$^{-1}$.}
\end{multline} 
Note that the obtained velocity does not depend on the magnetic-field strength and temperature. The corresponding evolution timescale $\tau_1$ turns out to be 
$C\sim 10$
times smaller than the timescale $\tau$ given by equation~(\ref{tau1})
\begin{equation}
\tau_1 \sim \frac{L}{V_{\rm L}} \sim \frac{\tau}{C} \sim  
\frac{3 \times 10^8}{C}
\frac{L_6^2}{H_{\rm c1,15}}~{\rm yr}.
\label{tau3}
\end{equation}
It is interesting, that the timescale (\ref{tau3}) is comparable to the 
typical pulsar age.

We arrive at the following possible picture of the evolution of arbitrary magnetic field in pulsars. At the first stage, an arbitrary magnetic field in a pulsar reconfigures itself to meet the condition (\ref{cond}). In an idealised setup 
($T$ is sufficiently low during the whole evolution process)
this reconfiguration occurs on a timescale $\sim \tau_{\rm init}$. In reality, however, a newly-born NS is hot, so that the condition (\ref{cond}) will be almost reached on the timescale of NS cooling, $\tau_\text{cool} \sim 10^4-10^6\,$yrs (e.g., \citealt{ppp15}). 
The latter timescale
is much longer than $\tau_{\rm init}$ 
and corresponds to a timescale of turning an NS into a cold pulsar studied here.

After that the evolution slows down dramatically, and at the second stage it proceeds on the timescale $\sim \tau_1$. During this time the next largest `driving' term in equation~(\ref{umuMinue}), 
proportional to
$\propto {\pmb e}_B \times {\pmb \nabla }{d}_{\rm e \mu}^\infty$, decreases 
\footnote{The third term in Eq.\ (\ref{umuMinue}) will still decrease at this stage, getting closer and closer to zero.}, 
approaching ${\pmb e}_B \times {\pmb \nabla }{d}_{\rm e \mu}^\infty \approx 0$
\footnote{Note that in the case of axisymmetric poloidal magnetic field the term ${\pmb e}_B \times {\pmb \nabla }{d}_{\rm e \mu}^\infty$ does not contribute to $V_{\rm L}$. Thus, for an axisymmetric problem it would be sufficient for the magnetic field to evolve to purely poloidal configuration during the second stage, vanishing of ${\pmb e}_B \times {\pmb \nabla }{d}_{\rm e \mu}^\infty$ is not necessary. Keeping in mind that purely poloidal magnetic field, at least in non-superconducting NSs, is unstable (\citealt*{fr77}), we could expect rearrangement of the magnetic field on a timescale $\sim \tau_1$. To examine this possibility one needs to perform a detailed modelling of B-evolution.}.
Together with the equation (\ref{cond}), this condition implies ${\pmb \nabla} {d}_{\rm e\mu}^\infty\approx 0$ 
(more precisely, 
${\pmb \nabla} {d}_{\rm e\mu}^\infty$ should be strongly suppressed,  
$|{\pmb \nabla} {d}_{\rm e\mu}^\infty|\sim (D_l/D_{l}') \mathcal{F}_{\rm b+t}/n_{\rm e}$,
i.e.,
become comparable to 
the second term in the r.h.s.\ of equation \ref{dtilde}).
In other words, the magnetic field becomes almost `barotropic', obeying the equation (see equation~\ref{tot}):
$n_{\rm p} {\pmb \nabla}\Delta \mu_\mu^\infty
={\pmb {\mathcal{F}}_{\rm b+t}},$
which, given that ${\pmb \nabla} d_{\rm e\mu}^\infty\approx 0$, actually coincides with the similar equation for npe-matter (see, e.g., \citealt{papm17b}). Note that, along with decreasing of the second and third terms in equation~(\ref{umuMinue}), the first term (and hence $\Delta {\pmb u}_{\rm \mu e}$ itself) is also automatically decreasing. Moreover, the contribution to $\partial {\pmb B}/\partial t$ from the second term in equation~(\ref{Vl}) decreases as well, and vanishes for `barotropic' field: 
${\pmb \nabla}\times\left(\frac{c {\pmb \nabla}\times {\pmb H}_{\rm c1}}{4 \pi e_{\rm p}n_{\rm p}}\times {\pmb B}\right)=c/e_{\rm p}{\pmb \nabla}\times ({\pmb {\mathcal F}}_{\rm b+t}/n_{\rm p})\approx c/e_{\rm p}{\pmb \nabla}\times ({\pmb \nabla}\Delta \mu_\mu^\infty)= 0$.
Then the fourth term in equation~(\ref{Vl}) comes into play, giving rise to the following estimate for the timescale of the subsequent slow B-evolution:
\begin{align}
\tau_2 \sim \frac{L}{V_{\rm L}}\sim \frac{ 4 \pi L^2}{\mathcal{K}_2 H_{\rm c1} B}
\sim \frac{\mathcal{D}_l'}{\mathcal{D}_l}\, \tau \sim 100 \tau, 
\label{tau4}
\end{align}
where $\tau$ is specified in equation~(\ref{tau1}).

In this section we adopted a model of homogeneous magnetic field. However, for estimates made above [in particular, for the timescales (\ref{tau1}), (\ref{tau2}), (\ref{tau3}), and (\ref{tau4})] the homogeneity of the field was used only when calculating the numerical coefficient $C \sim 10$, relating ${u}_{\rm e}^{({\rm pol})}$ and $\Delta u_{\rm \mu e}^{({\rm pol})}$. We expect that for the inhomogeneous magnetic field the coefficient $C$ will be even larger (correspondingly, the timescales $\tau_{\rm init}$ and $\tau_1$ will be smaller), because, as we already mentioned, ${u}_{\rm e}^{({\rm pol})}$ depends on high derivatives of the magnetic field, which can be very large (especially for the mixed poloidal-toroidal configurations of the magnetic field; see also a detailed discussion in \citealt{og18}). However, only accurate calculation may reveal the actual value of the coefficient $C$ for such field models.

\section{Summary and discussion}
\label{concl}

The problem of the magnetic field evolution in superfluid and superconducting NSs is addressed. It is shown that it can be studied within the framework developed by \cite{gko17,og18}. In this approach all particle velocities and chemical potential perturbations are calculated self-consistently provided that the magnetic field is specified. For illustration of the method we consider a simplified model of 
the NS core
composed of ${\rm npe\mu}$-matter, 
where
neutrons are strongly superfluid, and protons are strongly superconducting and form a type-II superconductor (a model of ordinary pulsar). All numerical results are obtained for a constant magnetic field $B=10^{12}$~G, which is not force-free in superconducting matter. We find (in agreement with \citealt{og18,crv20}, who considered normal strongly magnetised NSs) that the macroscopic particle velocities induced by the magnetic field are much larger (by a factor of $C\sim 10$) than the relative velocities. This finding contradicts a general belief that 
the bulk of NS matter can be treated as motionless to a good approximation.

Numerical results obtained in this paper allowed us to present the following 
qualitative picture of B-evolution in ${\rm npe\mu}$ cores of pulsars,
consisting of three stages:

(i) Magnetic field in a sufficiently cold pulsar has to meet the condition (\ref{cond}), that is imbalance of chemical potentials $d_{\rm e\mu}^\infty=\mu_{\rm e}^\infty-\mu_{\rm \mu}^\infty$ should be approximately constant along the magnetic field lines. Mathematically, this condition manifests itself in the fact that an arbitrary configuration of the magnetic field should evolve towards the configuration satisfying the condition (\ref{cond}) on an extremely short timescale $\sim \tau_{\rm init}$, see equation~(\ref{tau2}). In real pulsars this condition is achieved on the cooling timescale, 
$\tau_{\rm cool} \sim 10^4-10^6\,\rm yrs$ (see, e.g., \citealt{ppp15}),
during which
an NS 
approaches the low-temperature limit considered in the paper.
Our finding that the core  magnetic field initially evolves on the cooling timescale $\tau_{\rm cool}$ is in line with the results of \cite{ipt14,ip14,ip15,ip20}, who found similar timescale for the B-evolution
of young pulsars.

(ii) The evolution of the field, meeting the condition (\ref{cond}), proceeds on a timescale $\tau_1$ (see equation~\ref{tau3}), independent of the field 
strength
$B$ and temperature $T$. The latter timescale is comparable to the
typical pulsar age $\sim 10^7$~yr
\citep{ATNFref}\footnote{\texttt{www.atnf.csiro.au/people/pulsar/psrcat/}}.
This
probably indicates that the evolution of the core magnetic field plays an important role in the pulsar life and can
directly affect the observation properties of pulsars
(\citealt{gonthier04,popov10,gmvp14,gullon15,igoshev19}). The final result of this stage is an approximately barotropic magnetic field configuration (${\pmb \nabla}d_{\rm e\mu}^\infty\approx 0$) and very low particle velocities (both absolute and relative). 
Definitely, it would be very interesting to check if such `barotropic' magnetic-field configurations can be stable in superconducting NSs (\citealt{mbrsvl15,lj12}). If not, after a time interval $\sim \tau_1$, when the magnetic field reconfigures itself to satisfy the condition ${\pmb \nabla}d_{\rm e\mu}^\infty\approx 0$, the instability could set in and perturb the field out of `barotropic' state again. This process could be accompanied by a strong magnetic-field dissipation, leading to effective decay of the field on a timescale $\sim \tau_1$. See also the footnote 7, where another possibility, also potentially resulting in the instability onset 
after a time interval $\sim \tau_1$, is considered.

(iii) At the third stage particle velocities are already very low. As a result, the timescale $\tau_2$ of the subsequent B-evolution is very large (see equation~\ref{tau4}), exceeding or comparable to the lifetime of millisecond pulsars. In other words, once the magnetic field becomes `barotropic', its subsequent evolution in the stellar core is strongly suppressed. We come to conclusion that the magnetic field can hardly reach values typical to millisecond pulsars, $B\sim 10^8\,\rm G$, by means of the (exclusively) processes in the NS core (unless the instability comes into play, see above).

Our results imply that the magnetic field in the core should affect the observational manifestations of pulsars: It neither vanishes there nor `frozen', as it is often assumed in the literature. The details of the evolution can only be understood by performing self-consistent simulations of the B-evolution within the approach developed in this work.

In the end, let us briefly discuss few assumptions made in this study.
The first and foremost, we assumed that NS matter is sufficiently cold, so that most of neutrons and protons are paired. Conceptually, it is not a problem to extend our approach to the case of young (and warm) NSs. However, to do that, one needs to know the mutual friction parameters, describing interaction between the unpaired baryons and flux tubes, which are poorly known. We expect that the finite-temperature effects should slow down the B-evolution in the core significantly. In particular, according to \cite{og18}, the timescale of B-evolution in nonsuperfluid/nonsuperconducting star with the standard field $B\sim 10^{12}$~G and temperature $T\sim (3 \times 10^7-3\times 10^9)$~K is much larger than the corresponding cooling timescale. That is, apparently, the magnetic field does not change significantly due to the processes in the core before NS reaches temperatures 
at which the approximation of completely nonsuperfluid/nonsuperconducting matter is no longer valid.

The second important assumption is that we neglected interaction between the proton and neutron vortices in the course of NS evolution, by formally considering a non-rotating pulsar. Meanwhile, as \cite*{sbmt90,dcc93} proposed, this interaction may help to more effectively expel the magnetic field from the stellar 
core on a pulsar spin-down timescale (see, however, \citealt*{ga16}, arguing that this mechanism is inefficient).

\section*{Acknowledgments}

We thank A.~Reisenegger and F.~Castillo for discussions and S.B.~Popov for critical comments.
M.G. is partly supported by the Foundation for the Advancement of Theoretical Physics and mathematics `BASIS' [Grant No.\ 17-12-204-1] and by RFBR [Grant No.\ 19-52-12013]. D.O. acknowledges support from RFBR [Grant No.\ 19-52-12013]. Work of E.K. on the analysis of possible evolution scenarios has been supported by the Russian Science Foundation [Grant No.\ 19-12-00133].

\appendix

\section{Timescale of the magnetic field evolution in $npe$-matter}
\label{app}

Here we discuss the traditional estimate for the conservative timescale of the B-evolution, 
analogous to equation~(\ref{tau1}), but applicable to $npe$-matter.
Consider cold NS matter containing superconducting protons, superfluid neutrons, and electrons.
The Euler-like equations in this case are
\begin{align}
&
n_{\rm p} {\pmb \nabla} \mu_{\rm p}^\infty = 
{\pmb { \mathcal F}}_{\rm b+t}
-{\pmb {\mathcal F}}_{\rm pe}
+e_{\rm p} n_{\rm p} \, 
\left(
{\pmb E}+\frac{{\pmb V}_{\rm sp}}{c}\times {\pmb B}
\right),
&
\label{euler_pa}\\
&
n_{\rm n} {\pmb \nabla} \mu_{\rm n}^\infty = 0,
&
\label{euler_na}\\
&
n_{\rm e} {\pmb \nabla} \mu_{\rm e}^\infty = 
{\pmb {\mathcal F}}_{\rm pe}+
e_{\rm e} n_{\rm e} \, \left(
{\pmb E}+\frac{{\pmb u}_{\rm e}}{c}\times {\pmb B}
\right).
&
\label{euler_ea}
\end{align}
Summing up (\ref{euler_pa}) and (\ref{euler_ea}) 
and subtracting 
from the result
$n_{\rm p}/n_{\rm n}\times$(\ref{euler_na}), we get
\begin{align}
n_{\rm p} {\pmb \nabla}\Delta \mu_{\rm e}^\infty
={\pmb {\mathcal{F}}_{\rm b+t}},
\label{tota}
\end{align}
where $\Delta \mu_{\rm e}^\infty \equiv \mu_{\rm e}^\infty+\mu_{\rm p}^\infty-\mu_{\rm n}^\infty$.

We can use the same formula for the B-evolution timescale, 
$\tau = B/|{\pmb \nabla}\times ({\pmb V}_{\rm L}\times {\pmb B})|$, 
as in section~\ref{simple}, 
but now ${\pmb V}_{\rm L}$ equals
\begin{align}
&
{\pmb V}_{\rm L} \approx {\pmb V}_{\rm sp}+\left[1+\mathcal{O}\left(\frac{\mathcal{D}^2_{\rm e}}{\mathcal{D}'^{\,2}_{\rm e}}\right)\right]\frac{c {\pmb \nabla}\times {\pmb H}_{\rm c1}}{4 \pi e_{\rm p}n_{\rm p}}
+\mathcal{K}_2 
\frac{{\pmb \nabla}\times {\pmb H}_{\rm c1}}{4 \pi}  
\times {\pmb B},
&
\label{Vla}
\end{align}
where 
$\mathcal{K}_2 = \mathcal{D}_{\rm e}/\mathcal{D}'^{\,2}_{\rm e}$, 
and the term in square brackets is very close to 1 with small addition of the order of 
$\mathcal{D}_{\rm e}^2/\mathcal{D}'^{\,2}_{\rm e}$;
the actual form of this small term is not important for us here
\footnote{Similar coefficient is relevant for $npe\mu$ matter as well, 
but we did not introduce it in equation~(\ref{Vl}) since this small term was not important for us in the main text of the paper.}.
Following now the conventional way of estimating the conservative timescale 
(see, e.g., \citealt{gagl15,dg17,papm17b}),  
we skip ${\pmb V}_{\rm sp}$ 
\footnote{We emphasise once again that the approximation of motionless background is incorrect, 
but here we use it to follow the traditional consideration.}
and the last `dissipative' term, keeping only the second term in (\ref{Vla}), 
\begin{equation}
{\pmb V}_{\rm L} \approx \left[1+\mathcal{O}\left(\frac{\mathcal{D}^2_{\rm e}}{\mathcal{D}'^{\,2}_{\rm e}}\right)\right]\frac{c {\pmb \nabla}\times {\pmb H}_{\rm c1}}{4 \pi e_{\rm p}n_{\rm p}}.
\end{equation}
Using now the definition of ${\pmb {\mathcal F}}_{\rm b+t}$ and equation~(\ref{tota}) we get 
\begin{multline}
{\pmb \nabla}\times ({\pmb V}_{\rm L}\times {\pmb B}) = \frac{c}{e_{\rm p}} {\pmb \nabla}\times \left(\left[1+\mathcal{O}\left(\frac{\mathcal{D}^2_{\rm e}}{\mathcal{D}'^{\,2}_{\rm e}}\right)\right]{\pmb {\mathcal F}}_{\rm b+t}/n_{\rm p}\right) \\
=\frac{c}{e_{\rm p}} {\pmb \nabla}\times \left(\left[1+\mathcal{O}\left(\frac{\mathcal{D}^2_{\rm e}}{\mathcal{D}'^{\,2}_{\rm e}}\right)\right]{\pmb \nabla}\Delta \mu_{\rm e}^\infty\right),
\end{multline}
and hence the conservative timescale of B-evolution is
\begin{eqnarray}
\tau \sim 3 \times 10^8 \, \frac{L_6^2}{ H_{\rm c1, 15}}\,\,\frac{n_{\rm p}}{0.05\,\rm fm^{-3}}\, \times\mathcal{O}\left(\frac{\mathcal{D}'^2_{\rm e}}{\mathcal{D}^{\,2}_{\rm e}}\right) \ {\rm yr}.
\end{eqnarray}
where we estimated ${\pmb \nabla }\Delta \mu_{\rm e}^\infty$ as
$\sim H_{\rm c1}B/(4 \pi n_{\rm e}L)$ (see equation~\ref{tota}).
We see that B-evolution is suppressed by a factor 
$\mathcal{O}\left(\frac{\mathcal{D}'^2_{\rm e}}{\mathcal{D}^{\,2}_{\rm e}}\right)\sim 10^{4}$ 
(see \citealt{gusakov19} and figure~\ref{Fig:D}) 
in comparison to the case of $npe\mu$-matter.
This result 
agrees with \cite{gagl15,papm17b}.

We should note, however, that \cite{papm17b} 
came to this conclusion in a somewhat inconsistent way. 
Instead of equation~(\ref{euler_na}) for superfluid neutrons, 
\cite{papm17b} used an equation
with the right-hand side containing the terms responsible for the interaction of neutrons with other particle species, 
see equation (29) of that paper. 
We emphasise that this equation is
valid only if neutrons are completely nonsuperfluid 
(i.e., when $T>T_{\rm cn}$, where $T_{\rm cn}$ is the neutron critical temperature);
it cannot be used at $T<T_{\rm cn}$, in particular, in the limit of vanishing stellar temperature $T$.
The reason for that is the potentiality condition 
for the velocity of superfluid neutrons (\citealt{kg17}): Once $T$ falls below $T_{\rm cn}$, 
${\pmb \nabla} \mu_{\rm n}^\infty$ must vanish. 
Actually, the system (27)--(29) of \cite{papm17b} 
describes NS matter composed of superconducting protons, non-superfluid neutrons, and electrons. 
But in this case 
the timescale of conservative B-evolution is given by Eq.\ (\ref{tau1}), 
as one can easily check following the consideration of Sec.\ \ref{simple} and this appendix.
So why did \cite{papm17b} 
find
that B-evolution should be suppressed by a factor of $\mathcal{O}\left(\mathcal{D}'^2_{\rm e}/\mathcal{D}^{\,2}_{\rm e}\right)$? 
To answer this question, for the sake of simplicity, 
let us assume that $T\ll T_{\rm cp}$ and protons do not scatter off the other particle species 
($F_{\rm pn}=0$ in notations of \cite{papm17b}; 
here and below we use the notations of that reference).
In this case $F_{\rm en}=-n_{\rm n}{\pmb \nabla }\hat{\mu}_{\rm n}^\infty$, 
see equation (29) of \cite{papm17b}. 
In turn, $n_{\rm n}{\pmb \nabla }\hat{\mu}_{\rm n}^\infty$ is
set by equation (30) of \cite{papm17b}, once the magnetic field is specified, 
and is of the order of the tension/buoyancy term, $\pmb T$. 
As a result, the term ${\pmb F}_{\rm en}/n_c\sim {\pmb T}/n_c$ 
in equation (33) of \cite{papm17b} is the leading one. 
However, \cite{papm17b} ignored this (most important) term 
and incorrectly found that the conservative magnetic field timescale is suppressed.


\begin{thebibliography}{}
\makeatletter
\relax
\def\mn@urlcharsother{\let\do\@makeother \do\$\do\&\do\#\do\^\do\_\do\%\do\~}
\def\mn@doi{\begingroup\mn@urlcharsother \@ifnextchar [ {\mn@doi@}
  {\mn@doi@[]}}
\def\mn@doi@[#1]#2{\def\@tempa{#1}\ifx\@tempa\@empty \href
  {http://dx.doi.org/#2} {doi:#2}\else \href {http://dx.doi.org/#2} {#1}\fi
  \endgroup}
\def\mn@eprint#1#2{\mn@eprint@#1:#2::\@nil}
\def\mn@eprint@arXiv#1{\href {http://arxiv.org/abs/#1} {{\tt arXiv:#1}}}
\def\mn@eprint@dblp#1{\href {http://dblp.uni-trier.de/rec/bibtex/#1.xml}
  {dblp:#1}}
\def\mn@eprint@#1:#2:#3:#4\@nil{\def\@tempa {#1}\def\@tempb {#2}\def\@tempc
  {#3}\ifx \@tempc \@empty \let \@tempc \@tempb \let \@tempb \@tempa \fi \ifx
  \@tempb \@empty \def\@tempb {arXiv}\fi \@ifundefined
  {mn@eprint@\@tempb}{\@tempb:\@tempc}{\expandafter \expandafter \csname
  mn@eprint@\@tempb\endcsname \expandafter{\@tempc}}}

\bibitem[\protect\citeauthoryear{{Beloborodov} \& {Li}}{{Beloborodov} \&
  {Li}}{2016}]{bl16}
{Beloborodov} A.~M.,  {Li} X.,  2016, \mn@doi [\apj]
  {10.3847/1538-4357/833/2/261}, \href
  {http://adsabs.harvard.edu/abs/2016ApJ...833..261B} {833, 261}

\bibitem[\protect\citeauthoryear{{Braithwaite} \& {Spruit}}{{Braithwaite} \&
  {Spruit}}{2004}]{bs04}
{Braithwaite} J.,  {Spruit} H.~C.,  2004, \mn@doi [\nat] {10.1038/nature02934},
  \href {https://ui.adsabs.harvard.edu/abs/2004Natur.431..819B} {431, 819}

\bibitem[\protect\citeauthoryear{{Bransgrove}, {Levin}  \&
  {Beloborodov}}{{Bransgrove} et~al.}{2018}]{blb18}
{Bransgrove} A.,  {Levin} Y.,   {Beloborodov} A.,  2018, \mn@doi [\mnras]
  {10.1093/mnras/stx2508}, \href
  {http://adsabs.harvard.edu/abs/2018MNRAS.473.2771B} {473, 2771}

\bibitem[\protect\citeauthoryear{{Castillo}, {Reisenegger}  \&
  {Valdivia}}{{Castillo} et~al.}{2017}]{crv17}
{Castillo} F.,  {Reisenegger} A.,   {Valdivia} J.~A.,  2017, \mn@doi [\mnras]
  {10.1093/mnras/stx1604}, \href
  {https://ui.adsabs.harvard.edu/abs/2017MNRAS.471..507C} {471, 507}

\bibitem[\protect\citeauthoryear{{Castillo}, {Reisenegger}  \&
  {Valdivia}}{{Castillo} et~al.}{2020}]{crv20}
{Castillo} F.,  {Reisenegger} A.,   {Valdivia} J.~A.,  2020, arXiv e-prints,
  \href {https://ui.adsabs.harvard.edu/abs/2020arXiv200613186C} {p.
  arXiv:2006.13186}

\bibitem[\protect\citeauthoryear{{Cruces}, {Reisenegger}  \& {Tauris}}{{Cruces}
  et~al.}{2019}]{crt19}
{Cruces} M.,  {Reisenegger} A.,   {Tauris} T.~M.,  2019, \mn@doi [\mnras]
  {10.1093/mnras/stz2701}, \href
  {https://ui.adsabs.harvard.edu/abs/2019MNRAS.490.2013C} {490, 2013}

\bibitem[\protect\citeauthoryear{{Ding}, {Cheng}  \& {Chau}}{{Ding}
  et~al.}{1993}]{dcc93}
{Ding} K.~Y.,  {Cheng} K.~S.,   {Chau} H.~F.,  1993, \mn@doi [\apj]
  {10.1086/172577}, \href
  {https://ui.adsabs.harvard.edu/abs/1993ApJ...408..167D} {408, 167}

\bibitem[\protect\citeauthoryear{{Dommes} \& {Gusakov}}{{Dommes} \&
  {Gusakov}}{2017}]{dg17}
{Dommes} V.~A.,  {Gusakov} M.~E.,  2017, \mn@doi [\mnras]
  {10.1093/mnrasl/slx011}, \href
  {http://adsabs.harvard.edu/abs/2017MNRAS.467L.115D} {467, L115}

\bibitem[\protect\citeauthoryear{{Dommes}, {Gusakov}  \& {Shternin}}{{Dommes}
  et~al.}{2020}]{dgs20}
{Dommes} V.~A.,  {Gusakov} M.~E.,   {Shternin} P.~S.,  2020, arXiv e-prints,
  \href {https://ui.adsabs.harvard.edu/abs/2020arXiv200609840D} {p.
  arXiv:2006.09840}

\bibitem[\protect\citeauthoryear{{Elfritz}, {Pons}, {Rea}, {Glampedakis}  \&
  {Vigan{\`o}}}{{Elfritz} et~al.}{2016}]{eprgv16}
{Elfritz} J.~G.,  {Pons} J.~A.,  {Rea} N.,  {Glampedakis} K.,   {Vigan{\`o}}
  D.,  2016, \mn@doi [\mnras] {10.1093/mnras/stv2963}, \href
  {http://adsabs.harvard.edu/abs/2016MNRAS.456.4461E} {456, 4461}

\bibitem[\protect\citeauthoryear{{Flowers} \& {Ruderman}}{{Flowers} \&
  {Ruderman}}{1977}]{fr77}
{Flowers} E.,  {Ruderman} M.~A.,  1977, \mn@doi [\apj] {10.1086/155359}, \href
  {https://ui.adsabs.harvard.edu/abs/1977ApJ...215..302F} {215, 302}

\bibitem[\protect\citeauthoryear{{Glampedakis}, {Andersson}  \&
  {Samuelsson}}{{Glampedakis} et~al.}{2011a}]{gas11}
{Glampedakis} K.,  {Andersson} N.,   {Samuelsson} L.,  2011a, \mn@doi [\mnras]
  {10.1111/j.1365-2966.2010.17484.x}, \href
  {http://adsabs.harvard.edu/abs/2011MNRAS.410..805G} {410, 805}

\bibitem[\protect\citeauthoryear{{Glampedakis}, {Jones}  \&
  {Samuelsson}}{{Glampedakis} et~al.}{2011b}]{gjs11}
{Glampedakis} K.,  {Jones} D.~I.,   {Samuelsson} L.,  2011b, \mn@doi [\mnras]
  {10.1111/j.1365-2966.2011.18278.x}, \href
  {http://adsabs.harvard.edu/abs/2011MNRAS.413.2021G} {413, 2021}

\bibitem[\protect\citeauthoryear{{Goldreich} \& {Reisenegger}}{{Goldreich} \&
  {Reisenegger}}{1992}]{gr92}
{Goldreich} P.,  {Reisenegger} A.,  1992, \mn@doi [\apj] {10.1086/171646},
  \href {http://adsabs.harvard.edu/abs/1992ApJ...395..250G} {395, 250}

\bibitem[\protect\citeauthoryear{{Gonthier}, {Van Guilder}  \&
  {Harding}}{{Gonthier} et~al.}{2004}]{gonthier04}
{Gonthier} P.~L.,  {Van Guilder} R.,   {Harding} A.~K.,  2004, \mn@doi [\apj]
  {10.1086/382070}, \href
  {https://ui.adsabs.harvard.edu/abs/2004ApJ...604..775G} {604, 775}

\bibitem[\protect\citeauthoryear{{Gourgouliatos} \& {Cumming}}{{Gourgouliatos}
  \& {Cumming}}{2014}]{gc14}
{Gourgouliatos} K.~N.,  {Cumming} A.,  2014, \mn@doi [Physical Review Letters]
  {10.1103/PhysRevLett.112.171101}, \href
  {http://adsabs.harvard.edu/abs/2014PhRvL.112q1101G} {112, 171101}

\bibitem[\protect\citeauthoryear{{Gourgouliatos} \&
  {Hollerbach}}{{Gourgouliatos} \& {Hollerbach}}{2018}]{gh18}
{Gourgouliatos} K.~N.,  {Hollerbach} R.,  2018, \mn@doi [\apj]
  {10.3847/1538-4357/aa9d93}, \href
  {https://ui.adsabs.harvard.edu/abs/2018ApJ...852...21G} {852, 21}

\bibitem[\protect\citeauthoryear{{Gourgouliatos} \& {Pons}}{{Gourgouliatos} \&
  {Pons}}{2020}]{gp20}
{Gourgouliatos} K.~N.,  {Pons} J.~A.,  2020, arXiv e-prints, \href
  {https://ui.adsabs.harvard.edu/abs/2020arXiv200103335G} {p. arXiv:2001.03335}

\bibitem[\protect\citeauthoryear{{Gourgouliatos}, {Cumming}, {Reisenegger},
  {Armaza}, {Lyutikov}  \& {Valdivia}}{{Gourgouliatos}
  et~al.}{2013}]{gcr_etal13}
{Gourgouliatos} K.~N.,  {Cumming} A.,  {Reisenegger} A.,  {Armaza} C.,
  {Lyutikov} M.,   {Valdivia} J.~A.,  2013, \mn@doi [\mnras]
  {10.1093/mnras/stt1195}, \href
  {http://adsabs.harvard.edu/abs/2013MNRAS.434.2480G} {434, 2480}

\bibitem[\protect\citeauthoryear{{Gourgouliatos}, {Wood}  \&
  {Hollerbach}}{{Gourgouliatos} et~al.}{2016}]{gwh16}
{Gourgouliatos} K.~N.,  {Wood} T.~S.,   {Hollerbach} R.,  2016, \mn@doi
  [Proceedings of the National Academy of Science] {10.1073/pnas.1522363113},
  \href {http://adsabs.harvard.edu/abs/2016PNAS..113.3944G} {113, 3944}

\bibitem[\protect\citeauthoryear{{Graber}, {Andersson}, {Glampedakis}  \&
  {Lander}}{{Graber} et~al.}{2015}]{gagl15}
{Graber} V.,  {Andersson} N.,  {Glampedakis} K.,   {Lander} S.~K.,  2015,
  \mn@doi [\mnras] {10.1093/mnras/stv1648}, \href
  {http://adsabs.harvard.edu/abs/2015MNRAS.453..671G} {453, 671}

\bibitem[\protect\citeauthoryear{{G{\"u}gercino{\u{g}}lu} \&
  {Alpar}}{{G{\"u}gercino{\u{g}}lu} \& {Alpar}}{2016}]{ga16}
{G{\"u}gercino{\u{g}}lu} E.,  {Alpar} M.~A.,  2016, \mn@doi [\mnras]
  {10.1093/mnras/stw1758}, \href
  {https://ui.adsabs.harvard.edu/abs/2016MNRAS.462.1453G} {462, 1453}

\bibitem[\protect\citeauthoryear{{Gull{\'o}n}, {Miralles}, {Vigan{\`o}}  \&
  {Pons}}{{Gull{\'o}n} et~al.}{2014}]{gmvp14}
{Gull{\'o}n} M.,  {Miralles} J.~A.,  {Vigan{\`o}} D.,   {Pons} J.~A.,  2014,
  \mn@doi [\mnras] {10.1093/mnras/stu1253}, \href
  {https://ui.adsabs.harvard.edu/abs/2014MNRAS.443.1891G} {443, 1891}

\bibitem[\protect\citeauthoryear{{Gull{\'o}n}, {Pons}, {Miralles},
  {Vigan{\`o}}, {Rea}  \& {Perna}}{{Gull{\'o}n} et~al.}{2015}]{gullon15}
{Gull{\'o}n} M.,  {Pons} J.~A.,  {Miralles} J.~A.,  {Vigan{\`o}} D.,  {Rea} N.,
    {Perna} R.,  2015, \mn@doi [\mnras] {10.1093/mnras/stv1644}, \href
  {https://ui.adsabs.harvard.edu/abs/2015MNRAS.454..615G} {454, 615}

\bibitem[\protect\citeauthoryear{{Gusakov}}{{Gusakov}}{2019}]{gusakov19}
{Gusakov} M.~E.,  2019, \mn@doi [\mnras] {10.1093/mnras/stz657}, \href
  {https://ui.adsabs.harvard.edu/abs/2019MNRAS.485.4936G} {485, 4936}

\bibitem[\protect\citeauthoryear{{Gusakov} \& {Dommes}}{{Gusakov} \&
  {Dommes}}{2016}]{gd16}
{Gusakov} M.~E.,  {Dommes} V.~A.,  2016, \mn@doi [\prd]
  {10.1103/PhysRevD.94.083006}, \href
  {http://adsabs.harvard.edu/abs/2016PhRvD..94h3006G} {94, 083006}

\bibitem[\protect\citeauthoryear{Gusakov, Kantor  \& Ofengeim}{Gusakov
  et~al.}{2017}]{gko17}
Gusakov M.~E.,  Kantor E.~M.,   Ofengeim D.~D.,  2017, \mn@doi [Phys. Rev. D]
  {10.1103/PhysRevD.96.103012}, 96, 103012

\bibitem[\protect\citeauthoryear{{Harding}}{{Harding}}{2013}]{harding13}
{Harding} A.~K.,  2013, \mn@doi [Frontiers of Physics]
  {10.1007/s11467-013-0285-0}, \href
  {https://ui.adsabs.harvard.edu/abs/2013FrPhy...8..679H} {8, 679}

\bibitem[\protect\citeauthoryear{{Heiselberg} \& {Hjorth-Jensen}}{{Heiselberg}
  \& {Hjorth-Jensen}}{1999}]{hhj99}
{Heiselberg} H.,  {Hjorth-Jensen} M.,  1999, \mn@doi [\apjl] {10.1086/312321},
  \href {http://adsabs.harvard.edu/abs/1999ApJ...525L..45H} {525, L45}

\bibitem[\protect\citeauthoryear{{Henriksson} \& {Wasserman}}{{Henriksson} \&
  {Wasserman}}{2013}]{hw13}
{Henriksson} K.~T.,  {Wasserman} I.,  2013, \mn@doi [\mnras]
  {10.1093/mnras/stt338}, \href
  {https://ui.adsabs.harvard.edu/abs/2013MNRAS.431.2986H} {431, 2986}

\bibitem[\protect\citeauthoryear{{Hollerbach} \& {R{\"u}diger}}{{Hollerbach} \&
  {R{\"u}diger}}{2004}]{hr04}
{Hollerbach} R.,  {R{\"u}diger} G.,  2004, \mn@doi [\mnras]
  {10.1111/j.1365-2966.2004.07307.x}, \href
  {http://adsabs.harvard.edu/abs/2004MNRAS.347.1273H} {347, 1273}

\bibitem[\protect\citeauthoryear{{Hoyos}, {Reisenegger}  \& {Valdivia}}{{Hoyos}
  et~al.}{2008}]{hrv08}
{Hoyos} J.,  {Reisenegger} A.,   {Valdivia} J.~A.,  2008, \mn@doi [\aap]
  {10.1051/0004-6361:200809466}, \href
  {http://adsabs.harvard.edu/abs/2008A26A...487..789H} {487, 789}

\bibitem[\protect\citeauthoryear{{Hoyos}, {Reisenegger}  \& {Valdivia}}{{Hoyos}
  et~al.}{2010}]{hrv10}
{Hoyos} J.~H.,  {Reisenegger} A.,   {Valdivia} J.~A.,  2010, \mn@doi [\mnras]
  {10.1111/j.1365-2966.2010.17237.x}, \href
  {http://adsabs.harvard.edu/abs/2010MNRAS.408.1730H} {408, 1730}

\bibitem[\protect\citeauthoryear{{Igoshev}}{{Igoshev}}{2019}]{igoshev19}
{Igoshev} A.~P.,  2019, \mn@doi [\mnras] {10.1093/mnras/sty2945}, \href
  {https://ui.adsabs.harvard.edu/abs/2019MNRAS.482.3415I} {482, 3415}

\bibitem[\protect\citeauthoryear{{Igoshev} \& {Popov}}{{Igoshev} \&
  {Popov}}{2014}]{ip14}
{Igoshev} A.~P.,  {Popov} S.~B.,  2014, \mn@doi [\mnras]
  {10.1093/mnras/stu1496}, \href
  {https://ui.adsabs.harvard.edu/abs/2014MNRAS.444.1066I} {444, 1066}

\bibitem[\protect\citeauthoryear{{Igoshev} \& {Popov}}{{Igoshev} \&
  {Popov}}{2015}]{ip15}
{Igoshev} A.~P.,  {Popov} S.~B.,  2015, \mn@doi [Astronomische Nachrichten]
  {10.1002/asna.201512232}, \href
  {https://ui.adsabs.harvard.edu/abs/2015AN....336..831I} {336, 831}

\bibitem[\protect\citeauthoryear{{Igoshev} \& {Popov}}{{Igoshev} \&
  {Popov}}{2020}]{ip20}
{Igoshev} A.~P.,  {Popov} S.~B.,  2020, arXiv e-prints, \href
  {https://ui.adsabs.harvard.edu/abs/2020arXiv200811737I} {p. arXiv:2008.11737}

\bibitem[\protect\citeauthoryear{{Igoshev}, {Popov}  \& {Turolla}}{{Igoshev}
  et~al.}{2014}]{ipt14}
{Igoshev} A.~P.,  {Popov} S.~B.,   {Turolla} R.,  2014, \mn@doi [Astronomische
  Nachrichten] {10.1002/asna.201312029}, \href
  {https://ui.adsabs.harvard.edu/abs/2014AN....335..262I} {335, 262}

\bibitem[\protect\citeauthoryear{{Jones}}{{Jones}}{1988}]{jones88}
{Jones} P.~B.,  1988, \mn@doi [\mnras] {10.1093/mnras/233.4.875}, \href
  {http://adsabs.harvard.edu/abs/1988MNRAS.233..875J} {233, 875}

\bibitem[\protect\citeauthoryear{{Jones}}{{Jones}}{1991}]{jones91}
{Jones} P.~B.,  1991, \mn@doi [\mnras] {10.1093/mnras/253.2.279}, \href
  {http://adsabs.harvard.edu/abs/1991MNRAS.253..279J} {253, 279}

\bibitem[\protect\citeauthoryear{{Jones}}{{Jones}}{2006}]{jones06}
{Jones} P.~B.,  2006, \mn@doi [\mnras] {10.1111/j.1365-2966.2005.09724.x},
  \href {http://adsabs.harvard.edu/abs/2006MNRAS.365..339J} {365, 339}

\bibitem[\protect\citeauthoryear{{Kantor} \& {Gusakov}}{{Kantor} \&
  {Gusakov}}{2018}]{kg17}
{Kantor} E.~M.,  {Gusakov} M.~E.,  2018, \mn@doi [\mnras]
  {10.1093/mnras/stx2682}, \href
  {http://adsabs.harvard.edu/abs/2018MNRAS.473.4272K} {473, 4272}

\bibitem[\protect\citeauthoryear{{Kaspi}}{{Kaspi}}{2010}]{kaspi10}
{Kaspi} V.~M.,  2010, \mn@doi [Proceedings of the National Academy of Science]
  {10.1073/pnas.1000812107}, \href
  {http://adsabs.harvard.edu/abs/2010PNAS..107.7147K} {107, 7147}

\bibitem[\protect\citeauthoryear{{Kaspi} \& {Kramer}}{{Kaspi} \&
  {Kramer}}{2016}]{km16}
{Kaspi} V.~M.,  {Kramer} M.,  2016, arXiv e-prints, \href
  {https://ui.adsabs.harvard.edu/abs/2016arXiv160207738K} {p. arXiv:1602.07738}

\bibitem[\protect\citeauthoryear{{Kojima} \& {Kisaka}}{{Kojima} \&
  {Kisaka}}{2012}]{kk12}
{Kojima} Y.,  {Kisaka} S.,  2012, \mn@doi [\mnras]
  {10.1111/j.1365-2966.2012.20509.x}, \href
  {https://ui.adsabs.harvard.edu/abs/2012MNRAS.421.2722K} {421, 2722}

\bibitem[\protect\citeauthoryear{{Kojima} \& {Suzuki}}{{Kojima} \&
  {Suzuki}}{2020}]{kk20}
{Kojima} Y.,  {Suzuki} K.,  2020, \mn@doi [\mnras] {10.1093/mnras/staa1045},
  \href {https://ui.adsabs.harvard.edu/abs/2020MNRAS.494.3790K} {494, 3790}

\bibitem[\protect\citeauthoryear{{Konenkov} \& {Geppert}}{{Konenkov} \&
  {Geppert}}{2000}]{kg00}
{Konenkov} D.,  {Geppert} U.,  2000, \mn@doi [\mnras]
  {10.1046/j.1365-8711.2000.03188.x}, \href
  {http://adsabs.harvard.edu/abs/2000MNRAS.313...66K} {313, 66}

\bibitem[\protect\citeauthoryear{{Konenkov} \& {Geppert}}{{Konenkov} \&
  {Geppert}}{2001}]{kg01}
{Konenkov} D.,  {Geppert} U.,  2001, \mn@doi [\mnras]
  {10.1046/j.1365-8711.2001.04469.x}, \href
  {http://adsabs.harvard.edu/abs/2001MNRAS.325..426K} {325, 426}

\bibitem[\protect\citeauthoryear{{Landau} \& {Lifshitz}}{{Landau} \&
  {Lifshitz}}{1980}]{ll80}
{Landau} L.~D.,  {Lifshitz} E.~M.,  1980, {Statistical physics. Pt.2}.
Pergamon Press, Oxford

\bibitem[\protect\citeauthoryear{{Lander}}{{Lander}}{2013}]{lander13}
{Lander} S.~K.,  2013, \mn@doi [\prl] {10.1103/PhysRevLett.110.071101}, \href
  {https://ui.adsabs.harvard.edu/abs/2013PhRvL.110g1101L} {110, 071101}

\bibitem[\protect\citeauthoryear{{Lander} \& {Jones}}{{Lander} \&
  {Jones}}{2012}]{lj12}
{Lander} S.~K.,  {Jones} D.~I.,  2012, \mn@doi [\mnras]
  {10.1111/j.1365-2966.2012.21213.x}, \href
  {https://ui.adsabs.harvard.edu/abs/2012MNRAS.424..482L} {424, 482}

\bibitem[\protect\citeauthoryear{{Manchester}, {Hobbs}, {Teoh}  \&
  {Hobbs}}{{Manchester} et~al.}{2005}]{ATNFref}
{Manchester} R.~N.,  {Hobbs} G.~B.,  {Teoh} A.,   {Hobbs} M.,  2005, \mn@doi
  [\aj] {10.1086/428488}, \href
  {https://ui.adsabs.harvard.edu/abs/2005AJ....129.1993M} {129, 1993}

\bibitem[\protect\citeauthoryear{{Mendell}}{{Mendell}}{1991}]{mendell91a}
{Mendell} G.,  1991, \mn@doi [\apj] {10.1086/170609}, \href
  {http://adsabs.harvard.edu/abs/1991ApJ...380..515M} {380, 515}

\bibitem[\protect\citeauthoryear{{Mitchell}, {Braithwaite}, {Reisenegger},
  {Spruit}, {Valdivia}  \& {Langer}}{{Mitchell} et~al.}{2015}]{mbrsvl15}
{Mitchell} J.~P.,  {Braithwaite} J.,  {Reisenegger} A.,  {Spruit} H.,
  {Valdivia} J.~A.,   {Langer} N.,  2015, \mn@doi [\mnras]
  {10.1093/mnras/stu2514}, \href
  {https://ui.adsabs.harvard.edu/abs/2015MNRAS.447.1213M} {447, 1213}

\bibitem[\protect\citeauthoryear{{Ofengeim} \& {Gusakov}}{{Ofengeim} \&
  {Gusakov}}{2018}]{og18}
{Ofengeim} D.~D.,  {Gusakov} M.~E.,  2018, \mn@doi [\prd]
  {10.1103/PhysRevD.98.043007}, \href
  {http://adsabs.harvard.edu/abs/2018PhRvD..98d3007O} {98, 043007}

\bibitem[\protect\citeauthoryear{{Passamonti}, {Akg{\"u}n}, {Pons}  \&
  {Miralles}}{{Passamonti} et~al.}{2017a}]{papm17}
{Passamonti} A.,  {Akg{\"u}n} T.,  {Pons} J.~A.,   {Miralles} J.~A.,  2017a,
  \mn@doi [\mnras] {10.1093/mnras/stw2936}, \href
  {http://adsabs.harvard.edu/abs/2017MNRAS.465.3416P} {465, 3416}

\bibitem[\protect\citeauthoryear{{Passamonti}, {Akg{\"u}n}, {Pons}  \&
  {Miralles}}{{Passamonti} et~al.}{2017b}]{papm17b}
{Passamonti} A.,  {Akg{\"u}n} T.,  {Pons} J.~A.,   {Miralles} J.~A.,  2017b,
  \mn@doi [\mnras] {10.1093/mnras/stx1192}, \href
  {https://ui.adsabs.harvard.edu/abs/2017MNRAS.469.4979P} {469, 4979}

\bibitem[\protect\citeauthoryear{{Pons} \& {Geppert}}{{Pons} \&
  {Geppert}}{2007}]{pg07}
{Pons} J.~A.,  {Geppert} U.,  2007, \mn@doi [\aap]
  {10.1051/0004-6361:20077456}, \href
  {https://ui.adsabs.harvard.edu/abs/2007A&A...470..303P} {470, 303}

\bibitem[\protect\citeauthoryear{{Pons} \& {Vigan{\`o}}}{{Pons} \&
  {Vigan{\`o}}}{2019}]{pv19}
{Pons} J.~A.,  {Vigan{\`o}} D.,  2019, \mn@doi [Living Reviews in Computational
  Astrophysics] {10.1007/s41115-019-0006-7}, \href
  {https://ui.adsabs.harvard.edu/abs/2019LRCA....5....3P} {5, 3}

\bibitem[\protect\citeauthoryear{{Popov}, {Pons}, {Miralles}, {Boldin}  \&
  {Posselt}}{{Popov} et~al.}{2010}]{popov10}
{Popov} S.~B.,  {Pons} J.~A.,  {Miralles} J.~A.,  {Boldin} P.~A.,   {Posselt}
  B.,  2010, \mn@doi [\mnras] {10.1111/j.1365-2966.2009.15850.x}, \href
  {https://ui.adsabs.harvard.edu/abs/2010MNRAS.401.2675P} {401, 2675}

\bibitem[\protect\citeauthoryear{{Potekhin}, {Pons}  \& {Page}}{{Potekhin}
  et~al.}{2015}]{ppp15}
{Potekhin} A.~Y.,  {Pons} J.~A.,   {Page} D.,  2015, \mn@doi [\ssr]
  {10.1007/s11214-015-0180-9}, \href
  {https://ui.adsabs.harvard.edu/abs/2015SSRv..191..239P} {191, 239}

\bibitem[\protect\citeauthoryear{{Rheinhardt} \& {Geppert}}{{Rheinhardt} \&
  {Geppert}}{2002}]{rg02}
{Rheinhardt} M.,  {Geppert} U.,  2002, \mn@doi [Physical Review Letters]
  {10.1103/PhysRevLett.88.101103}, \href
  {http://adsabs.harvard.edu/abs/2002PhRvL..88j1103R} {88, 101103}

\bibitem[\protect\citeauthoryear{{Shalybkov} \& {Urpin}}{{Shalybkov} \&
  {Urpin}}{1997}]{su97}
{Shalybkov} D.~A.,  {Urpin} V.~A.,  1997, \aap, \href
  {https://ui.adsabs.harvard.edu/abs/1997A&A...321..685S} {321, 685}

\bibitem[\protect\citeauthoryear{{Shternin}}{{Shternin}}{2008}]{shternin08}
{Shternin} P.~S.,  2008, \mn@doi [Soviet Journal of Experimental and
  Theoretical Physics] {10.1134/S1063776108080050}, \href
  {http://adsabs.harvard.edu/abs/2008JETP..107..212S} {107, 212}

\bibitem[\protect\citeauthoryear{{Sonin}}{{Sonin}}{1987}]{sonin87}
{Sonin} E.~B.,  1987, \mn@doi [Reviews of Modern Physics]
  {10.1103/RevModPhys.59.87}, \href
  {http://adsabs.harvard.edu/abs/1987RvMP...59...87S} {59, 87}

\bibitem[\protect\citeauthoryear{{Srinivasan}, {Bhattacharya}, {Muslimov}  \&
  {Tsygan}}{{Srinivasan} et~al.}{1990}]{sbmt90}
{Srinivasan} G.,  {Bhattacharya} D.,  {Muslimov} A.~G.,   {Tsygan} A.~J.,
  1990, Current Science, \href
  {https://ui.adsabs.harvard.edu/abs/1990CSci...59...31S} {59, 31}

\bibitem[\protect\citeauthoryear{{Urpin} \& {Shalybkov}}{{Urpin} \&
  {Shalybkov}}{1999}]{us99}
{Urpin} V.,  {Shalybkov} D.,  1999, \mn@doi [\mnras]
  {10.1046/j.1365-8711.1999.02341.x}, \href
  {http://adsabs.harvard.edu/abs/1999MNRAS.304..451U} {304, 451}

\bibitem[\protect\citeauthoryear{{Vigan{\`o}}, {Rea}, {Pons}, {Perna},
  {Aguilera}  \& {Miralles}}{{Vigan{\`o}} et~al.}{2013}]{vigano_etal13}
{Vigan{\`o}} D.,  {Rea} N.,  {Pons} J.~A.,  {Perna} R.,  {Aguilera} D.~N.,
  {Miralles} J.~A.,  2013, \mn@doi [\mnras] {10.1093/mnras/stt1008}, \href
  {http://adsabs.harvard.edu/abs/2013MNRAS.434..123V} {434, 123}

\makeatother
\end{thebibliography}

\label{lastpage}

\end{document}